\documentclass[journal=jpclcd,manuscript=article]{achemso}

\usepackage{color}
\usepackage[version=3]{mhchem} 
\usepackage[T1]{fontenc}       

\author{Dima Bolmatov}
\email{d.bolmatov@gmail.com}
\affiliation{National Synchrotron Light Source II, Brookhaven National Laboratory, Upton, NY 11973, USA}
\author{Mikhail Zhernenkov}
\affiliation{National Synchrotron Light Source II, Brookhaven National Laboratory, Upton, NY 11973, USA}
\author{Dmitry Zav'yalov}
\affiliation{Volgograd State Technical University, Volgograd, 400005 Russia}
\author{Stanislav Stoupin}
\affiliation{Advanced Photon Source, Argonne National Laboratory, Argonne, Illinois 60439, USA}
\author{Yong Q. Cai}
\affiliation{National Synchrotron Light Source II, Brookhaven National Laboratory, Upton, NY 11973, USA}
\author{Alessandro Cunsolo}
\email{acunsolo@bnl.gov}
\affiliation{National Synchrotron Light Source II, Brookhaven National Laboratory, Upton, NY 11973, USA}

\title[An \textsf{achemso} demo]
  {Revealing the Mechanism of the Viscous-to-Elastic Crossover in Liquids}


\begin{document}

\begin{tocentry}
\includegraphics[scale=0.20]{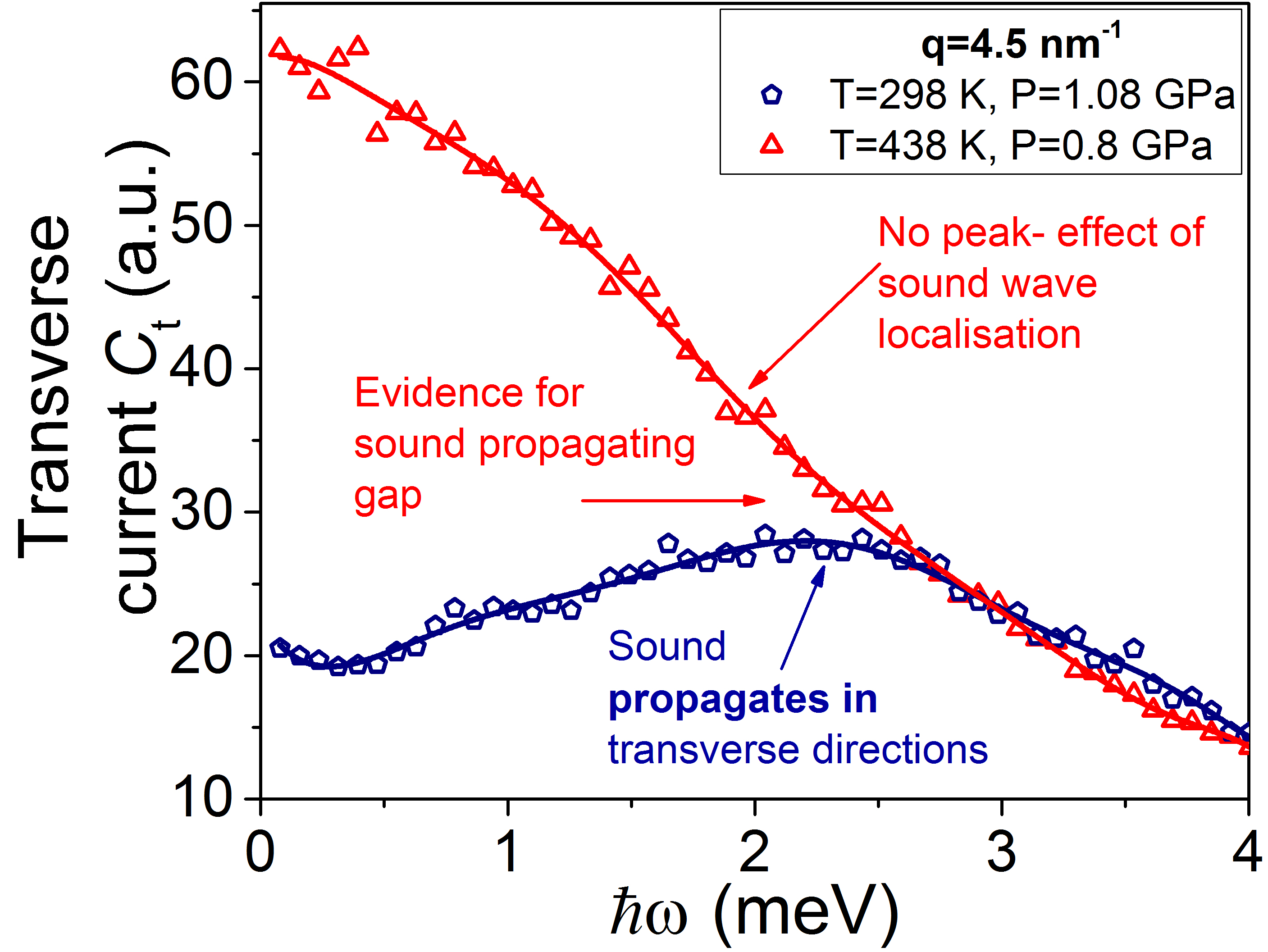}
\end{tocentry}

\begin{abstract}
  In this work we report on inelastic X-ray scattering experiments combined with the molecular dynamics simulations on deeply supercritical Ar. The presented results unveil the mechanism and regimes of sound propagation in the liquid matter, and provide a compelling evidence for the adiabatic-to-isothermal longitudinal sound propagation transition. We introduce a Hamiltonian predicting low-frequency transverse sound propagation gaps which is confirmed by experimental findings and molecular dynamics calculations. As a result, a universal link is established between the positive sound dispersion (PSD) phenomenon and the origin of transverse sound propagation revealing the viscous-to-elastic crossover in liquids. The PSD and transverse phononic excitations evolve consistently with theoretical predictions. Both can be considered as a universal fingerprint of the dynamic response of a liquid which is also observable in a sub-domain of supercritical phase. The simultaneous disappearance of both these effects at elevated temperatures is a manifestation of the Frenkel line. We expect that these findings will advance the current understanding of fluids under extreme thermodynamic conditions.   
\end{abstract}

The transition from the hydrodynamic to the single particle regime in the dynamics of monoatomic liquids is among the most fundamental topics to be investigated by spectroscopy experiments.  The collision event initiates the propagation of collective modes throughout the sample which is directly coupled with the spectrum of density fluctuations $S(q,\omega)$ measured by the experiment,   where $q$ is the wavenumber and $\omega$ is the frequency. This can be evidenced from the $q$ ($\omega$)  dependence of transport variables measured in an inelastic scattering experiment through their effect on $S(q,\omega)$. A noticeable case is the one of the sound velocity, which, at the departure from the hydrodynamic regime, undergoes a viscoelastic crossover with increasing $q$ from its liquid-like (viscous) to the solid-like (elastic) value. This viscoelastic phenomenon is usually referred to as the positive sound dispersion (PSD). It has extensively been  studied \cite{Balucani,Cunsolo_2001,Bencivenga_2007} and recently been connected with the aggregation phase and thermodynamic properties of a fluid \cite{Gorelli_2006,Simeoni_2010}. According to this interpretation, the PSD would be an universal fingerprint of the dynamic response of a liquid which is also observable
in a sub-domain of supercritical phase. More specifically, a thermodynamic boundary would demarcate the crossover from a liquid-like to a gas-like region in the supercritical phase and it would be characterized by a gradual disappearance of the PSD effect. This scenario is particularly fascinating because it implies a global reconsideration of the long-standing picture of the supercritical state as an intrinsically uniform phase. At the same time, more fundamental approaches such as the phonon theory of liquids predict a smooth transition of a supercritical liquid between a solid-like and a gas-like  response upon crossing a thermodynamic loci referred to as the Frenkel line \cite{bolsruni,bolnature,bolstr1,bolstr2}. The crossover between these two regions is  accompanied by the onset of characteristic solid-like features as, for instance, the ability in supporting the propagation of transverse phonon modes.

At present, no firm experimental evidences or theoretical results can be used in support of a universal link between the PSD effect and the origin of a transverse sound propagation in a liquid. This lack of knowledge  can  be overcome ideally by studying extreme supercritical conditions where PSD amplitude has been observed to disappear  experimentally \cite{Cunsolo_2001}, and the transverse sound propagation has been predicted to disappear theoretically \cite{bolsruni,bolnature,bolstr1,bolstr2}. Given these grounds, we decided to use inelastic X-ray scattering (IXS) technique and molecular dynamics (MD) simulations to determine the $S(q,\omega)$ of a simple monoatomic liquid (liquid Argon) in deeply supercritical conditions. 

\begin{figure*}[htp]
  \centering
 \begin{tabular}{cc}
    \includegraphics[width=150mm]{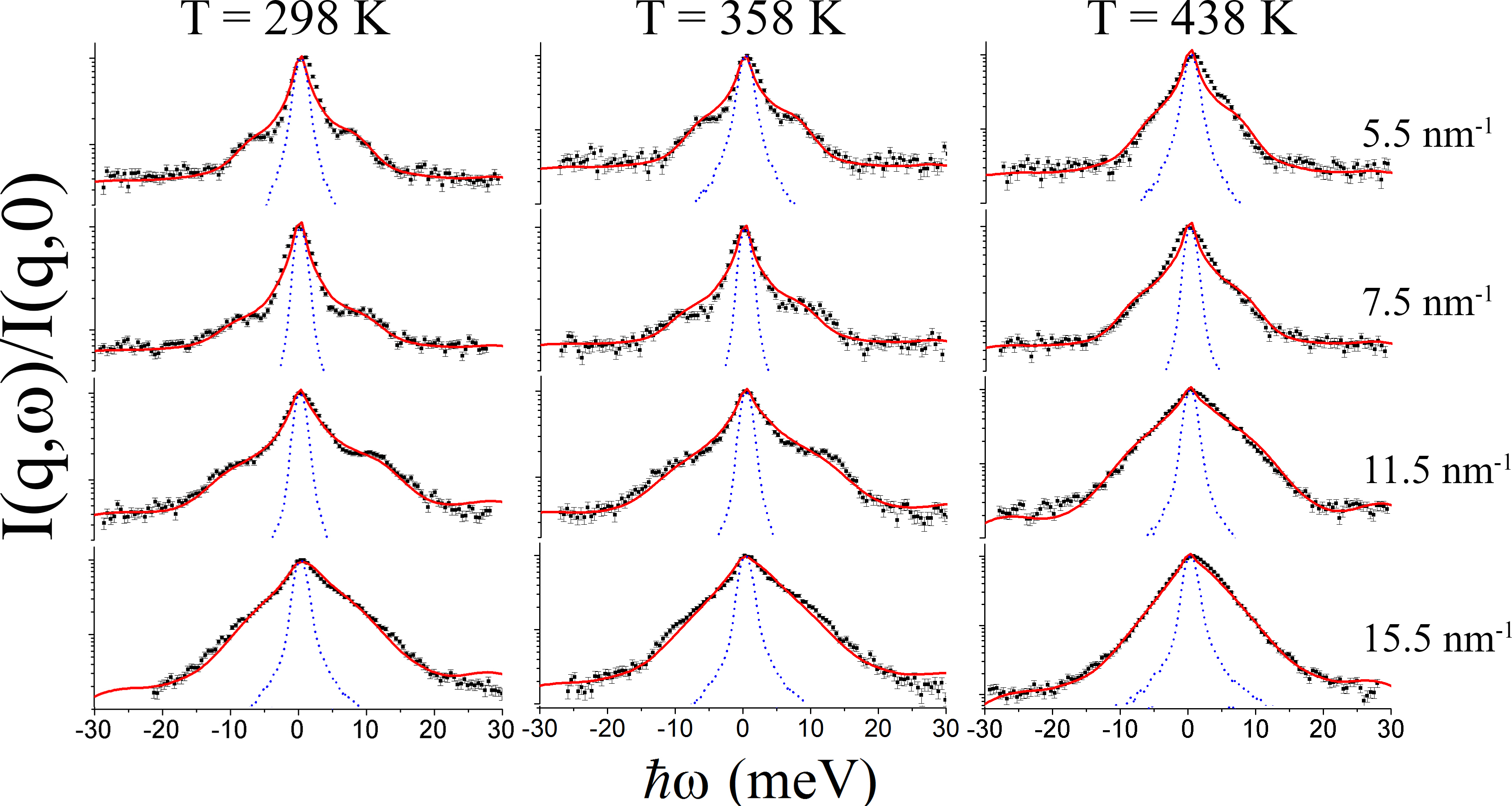}
  \end{tabular}
\caption{ IXS spectra of liquid Argon at different temperatures. Inelastic X-ray scattering spectra of Ar at T = 298 K, P = 1.08 GPa in the rigid liquid phase (Left panel), at T = 358 K, P = 0.9 GPa in the "mixed" phase at the Frenkel line (Central panel) and at T = 438 K, P = 0.8 GPa below the Frenkel line in the non-rigid fluid phase (Right panel). The experimental data (black squares) with error bars signifying one standard deviation are reported together with the MD simulation plus a sum of a delta function for the elastic component  (red solid curves). The MD curves are presented after convolution with the IXS spectrometer resolution function, which has almost Lorentzian shape (FWHM $\sim$2 meV) and is denoted by the dashed blue line. The corresponding $q$ values are shown to the right.}
\label{spectra}
\end{figure*} 

\begin{figure}[htp]
  \centering
 \begin{tabular}{cc}
    \includegraphics[width=75mm]{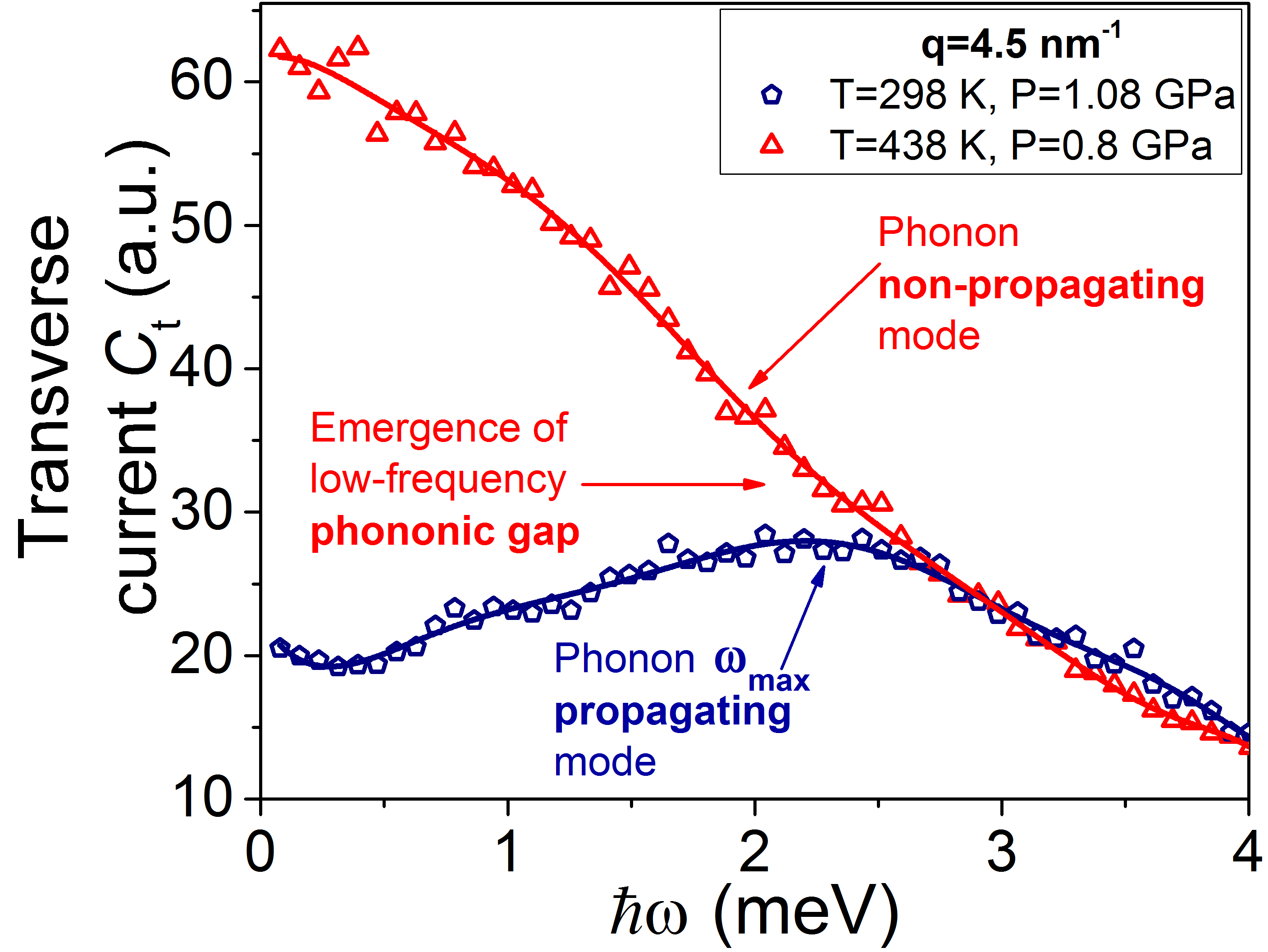}
  \end{tabular}
\caption{ Transverse current autocorrelation function $C_t$ above and below the Frenkel line. A clear peak (see the black line) indicates existence of transverse phonon propagating mode at T=298 K. The absence of the peak (see the red line, T=438) provides a compelling evidence for the emergence of the low-$q$ thermally-triggered phononic band gap, being predicted from Eq. (\ref{EHam}). Solid lines are guide to the eye only.}
\label{current}
\end{figure}

The evolution of the THz spectrum of liquid Ar beyond the hydrodynamic limit has been in the focus of a thorough scrutiny which include Brillouin light scattering \cite{Fleury}, inelastic neutron \cite{INS_Ar} (INS) and x-ray \cite{Simeoni_2010} scattering (IXS) measurements, as well as molecular dynamics simulations \cite{Barker,Levesque}.
One of the main aim of these works was to test the mode-coupling theory predictions for the line-shape evolution beyond the hydrodynamic limit \cite{Deschepper}. 
Despite such a comprehensive study, the evolution of the spectrum of Ar or any other monatomic system under extreme thermodynamic conditions still remains  unexplored. This subject bears important consequences for the ongoing effort in understanding the thermodynamic properties of disordered materials \cite{zalupa1,zalupa2,zalupa3,zalupa4}. To find a possible connection between the PSD phenomenon and the origin of transverse sound propagation one may consider the effective Hamiltonian \cite{bolsruni} where the problem of strong interactions is resolved from the outset
\begin{equation}
\label{EHam}
\begin{aligned}
H[\varphi_q] =\frac{1}{2}\sum_{0\leq\omega_q^{l,t,t}\leq\omega_{\rm D}}[\pi_q^{l}\pi_{-q}^{l}+\pi_q^{t}\pi_{-q}^{t}+\pi_q^{t}\pi_{-q}^{t}]+\\
\sum_{0\leq\omega_q^l\leq\omega_{\rm D}}\left[\frac{\omega_q^2}{2}\varphi^l_q\varphi^l_{-q}\right]+ \sum_{\omega_{\rm F}\leq\omega_q^{t,t}\leq\omega_{\rm D}}\left[\frac{\omega_q^2}{2}(\varphi^t_q\varphi^t_{-q}+\varphi^t_q\varphi^t_{-q})\right]
\end{aligned}
\end{equation}
where $q$ is a multiindex $\{q_l,q_t,q_t\}$, the collective canonical coordinates $\pi^\alpha_q$ and $\varphi_q^\alpha$ ($\alpha=l,t,t$) are defined as follows: $\varphi_q^\alpha= \sqrt{m} \sum_{j=1}^{N} e^{{\texttt i}L (j\cdot q) }x^\alpha_j$
and $\pi^\alpha_q = \dot{\varphi}_q^\alpha$. Here, $x_j^\alpha$  is space coordinate of an atom of a lattice sitting in a vertex labelled by the multiindex. $l$ and $t$ stand for the longitudinal and transverse directions of sound propagation, respectively. $\omega_{\rm D}$ is the Debye frequency, $\omega_{\rm F}$  defines a lower bound on the oscillation frequency of the atoms and can be derived from the viscosity $\eta$ and shear modulus $G_{\infty}$ of a fluid \cite{bolsrph,bolprb}
\begin{equation}
\omega_{\rm F}(T)=\frac{2\pi}{\tau(T)}=\frac{2\pi G_{\infty}}{\eta(T)}
\label{maxwell}
\end{equation}
where $\tau=\frac{\eta}{G_{\infty}}$ is the Maxwell's relation and $\tau$ is the relaxation time \cite{Frenkel}. This Hamiltonian predicts
the low-frequency transverse phononic bandgaps in a liquid spectrum ($\omega_{\rm F}\leq\omega_q^{t,t}\leq\omega_{\rm D}$, see the last term in Eq. (\ref{EHam}))
which is a result of a symmetry breaking due to interaction of phonons \cite{bolsruni}. Viscosity $\eta(T)$ drops down upon temperature rise resulting in the transverse phononic
band gap (the emergence of the sound propagation gap can be evidence below from Fig. (\ref{current})) growth  due to the Frenkel frequency $\omega_{\rm F}$ (see Eq. (\ref{maxwell})) increase \cite{bolsrph,bolprb}. An increase in temperature leads to the progressive disappearance of the high-frequency transverse phonon modes. This leads to $\omega_{\rm F}\xrightarrow{T}\omega_{\rm D}$ and, hence,  $c_V=\left(\frac{1}{N}\frac{\partial E}{\partial T}\right)_{\rm V}$ ($H\varphi=E\varphi$): 3$k_{\rm B}\xrightarrow{T}$ 2$k_{\rm B}$. $c_{V}=2k_{\rm B}$ (when $\omega_{\rm F}=\omega_{\rm D}$)  is the new thermodynamic limit \cite{bolsruni,bolnature,bolstr1,bolstr2}, dubbed here the Frenkel line thermodynamic limit. Further, we will present experimental and, then, MD simulations results to demonstrate the link between the evolution of the PSD phenomenon and the evolution of transverse sound propagation.  It is important to mention that even the first order phase transition such as the melting occurs at $c_V>$ 3$k_{\rm B}$, where the $c_V$=3$k_{\rm B}$ is the well-known value, namely the Dulong-Petit law. This means that the thermodynamic limit for solids is $c_V$=3$k_{\rm B}$ but the actual transition (the melting) takes place at $c_V\approx$ 3.1$k_{\rm B}$--3.5$k_{\rm B}$ due to anharmonic effects \cite{bolsrph}. The The Frenkel line is not a phase transition such as the melting line but is the continuous crossover which roughly begins at $c_V\approx$ 2.1$k_{\rm B}$--2.4$k_{\rm B}$ due to anharmonic effects and/or the effects influenced by the $\lambda$-like supercritical anomaly evidenced in the previous studies \cite{bolnature}. The similar behavior of the heat capacity was recently observed in a supercooled liquid \cite{prlgrigera}.

Selected spectra  normalized to the respective intensity for the liquid Argon, measured by inelastic X-ray scattering (IXS)  in a diamond anvil cell (DAC), are depicted in Fig. (\ref{spectra}) at three thermodynamic conditions: left panel (T=298 K, P=108 GPa), central panel (T=358 K, P=0.9 GPa) and right panel ( T=438 K, P=0.8 GPa). The high quality data clearly proves the existence of longitudinal phonon modes, which appears as peaks or shoulders in the lower $q$  range. The calculated dynamic structure factor from MD simulations  with a sum of a delta function for the elastic component was convoluted with the resolution function is in a good agreement with the experimental $S(q,\omega)$ (see Fig. (\ref{spectra})). Both simulated and measured spectra bear evidence of well defined inelastic shoulders whose position seems clearly increasing with $q$ at low/moderate exchanged momenta. It can also be noticed that lowest $q$ IXS spectra exhibit additional high frequency peaks in the upper extreme of the probed frequency range (see Fig. (\ref{spectra})), these peaks, not observable in the simulated spectra are ascribed to the phonons excitations dominating the scattering from the diamond DAC windows.

We performed high pressure/high temperature inelastic x-ray scattering measurements using a BX90 Diamond Anvil Cell \cite{kantor} at the IXS beam line 30-ID of the Advanced Photon Source (APS), Argonne National Laboratory. The DAC was used in combination with tungsten-carbide seats and full diamond anvils with a 500 $\mu$m culet size.  250 $\mu$m-thick rhenium gasket was  pre-indented to a thickness of about 90 $\mu$m. A hole with a diameter of about 220 $\mu$m was drilled in the middle of the pre-indented area. Conventional resistive heating was used to heat the sample. The $^{\rm 40}$Ar was loaded using a COMPRES/GSECARS gas-loading system at APS \cite{compress} up to initial pressure of 1.08 GPa. A ruby sphere was used for the pressure calibration \cite{sandeep}. The sample was measured at the photon energy of 23.7 keV. Following every temperature change, the DAC was allowed to equilibrate for, at least, 15 minutes before the IXS spectrum was collected.  The uncertainty related to the temperature measurement was about $\pm$ 5 K, which also resulted in a corresponding uncertainty of 0.1 GPa in the pressure determination from the ruby spectrum.  It is also important to mention that DAC may produce signal contamination in IXS analyzers due to diamond Bragg diffraction of the incident beam. During our experiment, the Bragg peaks were avoided by the proper choice of angular position of the DAC. Such angular position can be easily found at any given experimental condition since the divergence of the incident beam is about 1 mrad ($\sim$1 mm beam focused by Kirkpatrick-Baez mirrors with a focal distance of $\sim$1 m) and the minimum angular spacing of the Bragg reflections in diamond at 23.7 keV is $\sim$5 mrad.  It is possible that Bragg reflections of a diamond anvil cell have substantial angular acceptance (much larger than that of a perfect diamond crystal due to intrinsic and pressure-temperature induced strain and lattice tilts). In this case an elastic line in the particular analyzer channel can become contaminated. During reduction of raw IXS data no substantial difference in the height of the elastic line compared to the background IXS level was found for different analyzer channels, which confirms that the anvil cell was effectively acting as a window.

The MD simulation is intrinsically unable to probe the extremely long time response of the fluid ($\omega\cong 0$) mainly concentrating around the elastic position.  In order to perform a reliable comparison between measured and simulated spectra, an elastic $\propto \delta(\omega)-$function was added. The elastic term accounts for all parasitic intensity effects contributing to the measured elastic intensity. The MD lineshapes were corrected by a detailed balance factor which governs the positive and negative phonon energy intensities \cite{sette1}. The MD lineshapes combined with the elastic term were convoluted with the experimentally measured instrumental resolution functions. Furthermore, the experimental IXS background contribution was taken into account when comparing the MD simulations and the IXS data. The elastic intensity and background level were treated as variables whose optimized values provided the best fit to the experimental spectra.  The resulting MD lineshapes are reported as red solid lines in Figure (\ref{spectra}). To calculate dispersion relations and correlation functions, we have used LAMMPS simulation code to run a Lennard-Jones (LJ, $\varepsilon/k_{\rm B}$=119.8 K, $\sigma$=3.405) fluid fitted to Ar properties with 32678 atoms in the isothermal-isobaric (NPT) ensemble.
\begin{figure}[htp]
  \centering
 \begin{tabular}{cc}
    \includegraphics[width=75mm]{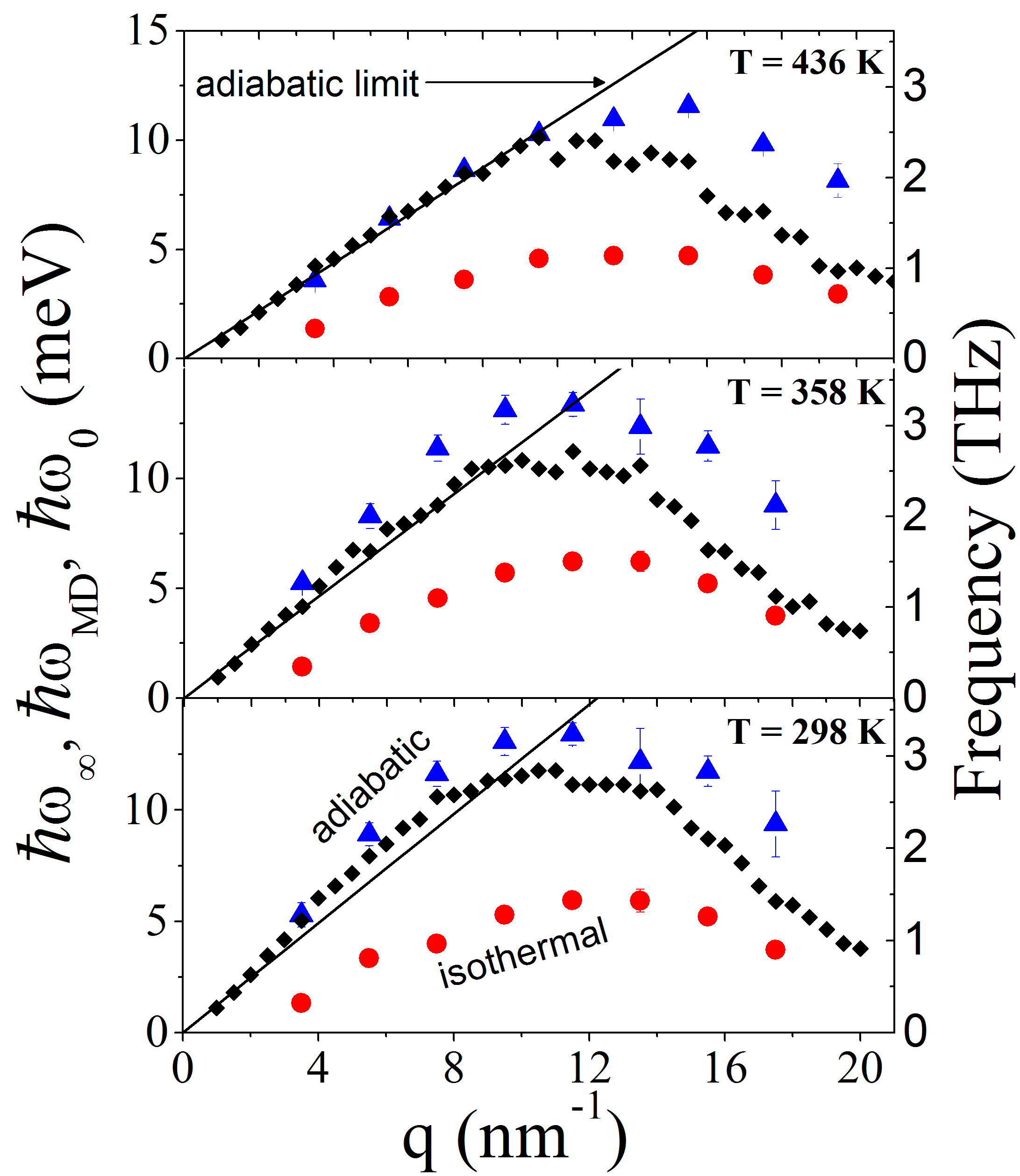}
  \end{tabular}
\caption{ Evidence for the adiabatic-to-isothermal longitudinal sound propagation crossover. Best fit dispersion curves $\omega_{\infty}$ (adiabatic, blue triangles) and $\omega_{0}$ (isothermal, red dots) are compared with the sound frequency $\omega_{MD}$ computed from maxima of longitudinal current autocorrelation functions (black dots) at the experimental conditions along the following P-T path ( T = 298 K, P = 1.08 GPa; T = 358 K, P = 0.9 GPa; T = 438 K, P = 0.8 GPa). The solid straight line represents adiabatic (or hydrodynamic) limit evaluated from the chemistry web-book\cite{NIST}.  }
\label{phonon}
\end{figure}

An  variable of exceptional interest for spectroscopy experiments can be considered using the following normalized correlation function
\begin{equation}
\Phi_q (t)=\frac{< \delta \rho^*_q (0) \delta \rho_q (t) >}{< \delta
\rho^*_q (0) \delta \rho_q (0) >}  \label{phi}
\end{equation}
with the variable $\delta \rho_q(t)$ being the $q$-component of the fluctuation of the microscopic number density $\rho (r,t)$. More specifically the spectroscopy experiment probes directly the Fourier transform of $\Phi_q (t)$ through the dynamic structure factor $S(q, \omega)$
\begin{equation}
S(q,\omega)=S(q) \int_{-\infty}^{\infty} dt e^{-i \omega t} \Phi_q(t),
\label{Sqw2}
\end{equation}
A continuous fraction representation for the density correlation function (see the textbook \cite{Balucani}) can be used
$\Phi_q(t)$ and its second-order truncation eventually yields
\begin{equation}
S(q,\omega)=\frac{2 v_o^2 q^2}{\omega} {\it Im}[\omega^2-\omega(q)
^2-i\omega m_q(\omega)]^{-1}.  \label{Sqom}
\end{equation}
where {\it Im} denotes the imaginary part and we have introduced the second memory function $m_q(\omega)$. Furthermore the fulfillment of the second sum rule of $S(q,\omega)$ in Eq. \ref{Sqom} yields
\begin{equation}
\omega_{q} =\sqrt{\frac{(qv_T)^2}{S(q)}} \label{mom2}
\end{equation}
where $v_T^2=k_{\rm B} T/M$ is the classical thermal speed defined in terms of the molecular mass $M$ and the Boltzmann constant $k_{\rm B}$. At this point the problem shifts from the search of a suitable model for $S(q,\omega)$ to  the one for the memory function $m_q(\omega)$. In the time domain this amounts to defining the most realistic and accurate description of the decay of the time dependent memory function. Here, we used a simplified expression for the memory function which consistently describes the high frequency spectrum of noble gases \cite{Levesque,Cunsolo_2001,Verbeni_2001} as well
as other simple fluids \cite{Cunsolo_2005}, specifically
\begin{equation}
K_l(q,t) =(\omega_{\infty}^2-\omega_0^2)e^{-t/\tau} \label{visco1}
\end{equation}
which is customarily referred as the Debye approximation, $ \omega_{\infty}$ is the elastic sound frequency, $\omega_0=\gamma\omega(q)=c_0 q$ is the viscous sound frequency,  while $c_0 (q)$ represents the finite-$q$ extension of the liquid-like sound velocity. The consistency of Eq. (\ref{visco1}) with the Navier-Stokes theory implies $c_0 = c_s$, where $c_s$ is the adiabatic sound speed, and one also has 
$c_s=\sqrt{\gamma} c_T$ with $c_T$ being the isothermal sound velocity. All parameters are $q$-dependent quantities, even though such a dependence is not explicitly mentioned in the notation. For non-metallic systems, at low-$q$  values  sound modes propagate adiabatically, i.e. without any thermal exchange with the propagating medium (adiabatic regime). The situation is reversed in high-$q$ isothermal regime, when the thermal exchange becomes more rapid than a period of acoustic oscillations. In this regime, a thermal equilibrium instantaneously reached between the acoustic wave and its propagation medium. The spectral lineshape was described with a model based on the memory function in Eq. (\ref{visco1}) in which the relevant best fit parameters $\omega_0$, $\omega_\infty$ and $\tau$ were used. 

\begin{figure}[htp]
  \centering
 \begin{tabular}{cc}
    \includegraphics[width=75mm]{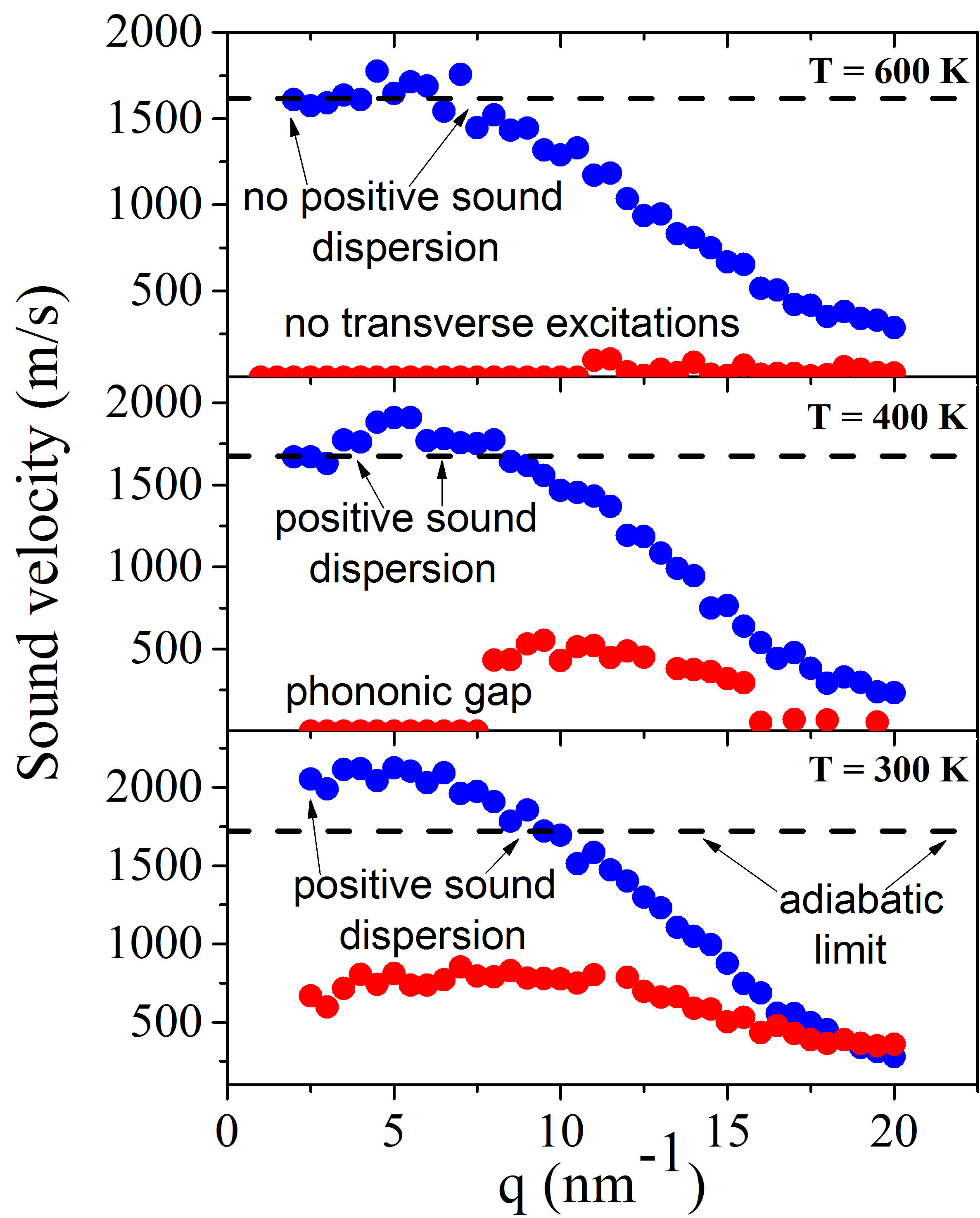}
  \end{tabular}
\caption{Longitudinal and transverse sound velocities. Longitudinal (blue dots) and transverse (red dots) sound velocities are reported as a function of $q$ as evaluated from the maxima positions of the current autocorrelation functions of corresponding polarizations at fixed pressure P=0.8 GPa. The dashed horizontal line represents the adiabatic (or hydrodynamic) limit.}
\label{sound}
\end{figure}

Best fit values of the viscoelastic parameters $\omega_{0}$ and $\omega_{\infty}$  are reported in Fig. (\ref{sound}) for the three pressure-temperature conditions probed in the experiment. They are compared with the sound frequency $\omega_{MD}$, derived from the position of the dominant peaks of the longitudinal current spectra and the hydrodynamic
linear dispersion  $c_0 q$ with $c_0$ being the adiabatic sound velocity derived from the NIST database \cite{NIST}.
 Some striking features readily emerges from the analysis of reported results, as detailed below. The most clear dispersion effects can be observed in the bottom graph of Fig. (\ref{sound}) (T=298 K), where at low/intermediate $q$-values both $\omega_{MD}$ and its elastic limit $\omega_{\infty}$ lay significantly above the hydrodynamic linear dispersion (solid straight line).
\begin{figure}[htp]
  \centering
 \begin{tabular}{cc}
    \includegraphics[width=60mm]{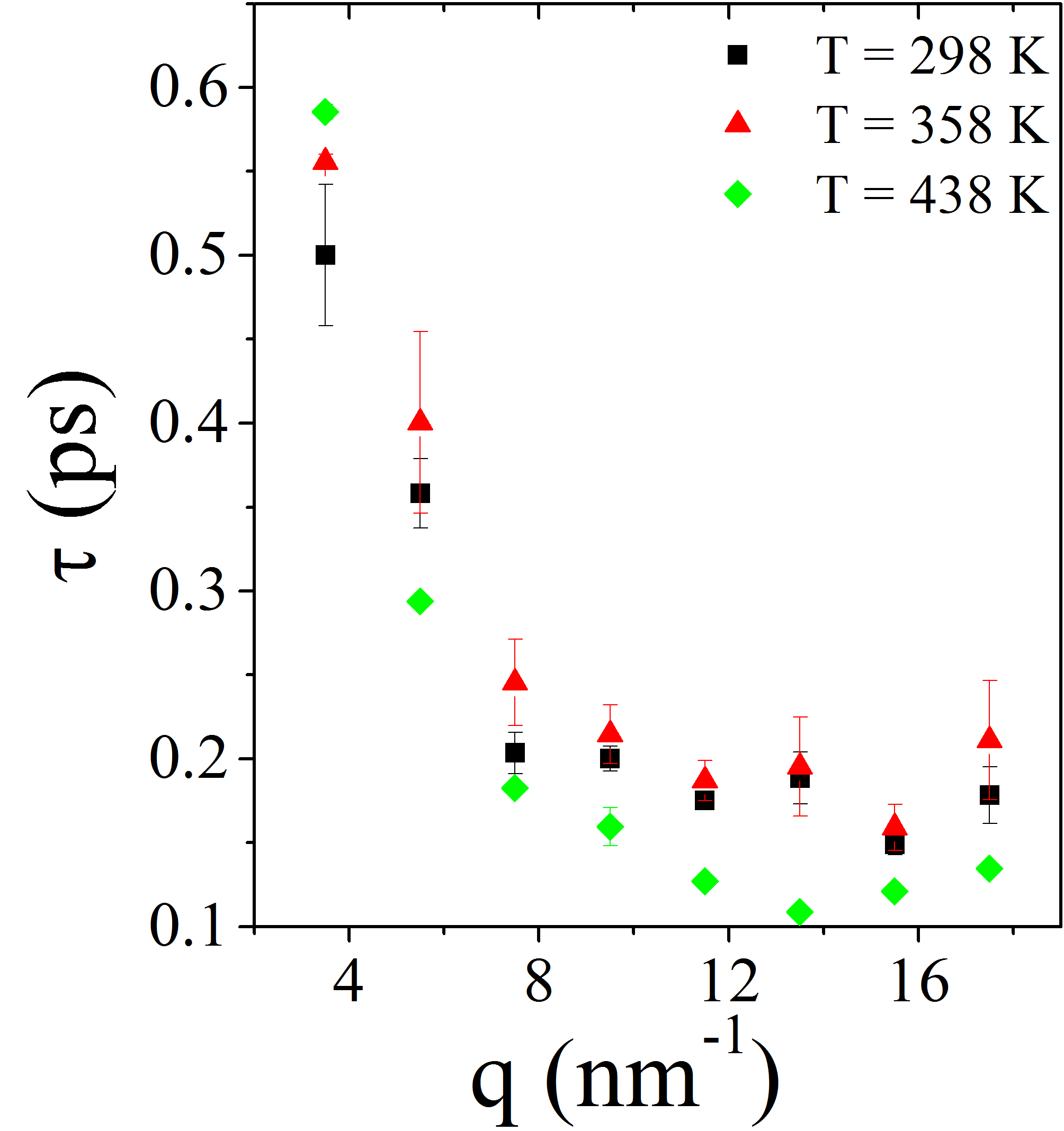}
  \end{tabular}
\caption{ Relaxation time at the experimental conditions. Best fit values of the relaxation time $\tau$ are reported as a function of $q$.}
\label{tau}
\end{figure}

Furthermore, $\omega_{MD}$ follows the expected more-than-linear low $q$-trend leading from the hydrodynamic linear dispersion up to the elastic value $\omega_{\infty}$ at intermediate $q$-values.
This is a clear manifestation of the PSD phenomenon mentioned in the introductory section. Most importantly, the amplitude of the PSD vanishes when the deepest supercritical conditions are reached (see upper plot). A similar trend is evident from the $\omega_{\infty}-c_0 q$ difference evolution. Overall, reported data pinpoint that all viscoelastic effects on the sound dispersion disappear when deeply supercritical conditions are approached, thereby confirming the findings of  previous IXS studies \cite{Cunsolo_2001,Gorelli_2006,Simeoni_2010}. For $q$-values higher than those covered by experimental data (17-20 $nm^{-1}$ range, see Fig. (\ref{sound})) $\omega_{MD}$ closely approaches the opposite viscous limit $\omega_{0}$.
Interestingly, in this $q$-range the specific heats ratio $\gamma \approx 1$ (see e.g. Ref. \cite{Levesque}), therefore, any distinction between adiabatic and isothermal sound propagation is lost.
This suggests that the sound propagation follows the expected adiabatic-to-isothermal crossover \cite{Bencivenga_2006}.
The $q$-dependence of relaxation time $\tau$ derived from the viscoelastic analysis of IXS spectra (see Fig. (\ref{spectra})) is shown in Fig. (\ref{tau}). One readily observes that a temperature increase clearly results in a sizable decrease of $\tau$ making faster local rearrangements in the fluid. One can notice that for the lowest q-point the trend is reversed because here the measurement of the relaxation time is resolution limited and the error of the $\tau$ is large. Furthermore, at low/intermediate $q$-values ($q< 8 nm^{-1}$) the $q$-dependence of $\tau$ becomes much pronounced which indicates a crossover from the hydrodynamic regime towards the cage oscillation one. The former regime is dominated by slow (ps scale) structural processes while the latter is governed by mutual atomic collisions.

The relaxation processes described above leave a clear fingerprint on the sound velocity which exhibits a $q$-increase usually referred to as the {\it positive sound dispersion}. This  phenomenon is a direct signature of the {\it viscous-to-elastic crossover}, which is a precursor of the shear propagation, observable only once the elastic limit is fully reached. The interconnection between these two phenomena is unambiguously demonstrated in Fig. \ref{sound}. Interestingly, despite several studies on the evolution of the PSD under supercritical conditions \cite{Cunsolo_2001,Gorelli_2006,Simeoni_2010} no firm connection has been established between this phenomenon and transverse sound propagation. Importantly, we observe that the decrease of PSD displays a continuous trend upon temperature increase. Its disappearance is accompanied by the parallel evanescence of the transverse propagating phonon mode and leads to the appearance of both low- and high-$q$ sound propagation gaps (see Fig. (\ref{sound})). We notice that at the highest $q$-values (T=300 K) the longitudinal and transverse sound velocities curves merge into each other. This implies that longitudinal and transverse currents spectra are dominated by the same mode, which is a clear manifestation of the so-called {\it longitudinal-transverse coupling} \cite{Sampoli}. Overall, the presented data clearly demonstrate that the PSD phenomenon is induced by the shear relaxation process, that is a relaxation process directly affecting the shear viscosity coefficient.

In conclusion, we have shown that the joint use of the inelastic X-ray scattering and the molecular dynamics simulations provides a consistent picture of the THz dynamic response of deeply supercritical Ar (see Fig. (\ref{spectra})). More specifically, we found that a Lennard-Jones model provides an accurate description of the lineshape over a broad range of thermodynamic conditions. The calculated transverse current autocorrelation function undergoes a crossover from a solid-like regime, where shear sound propagation is supported, to a gas-like one, where it is instead suppressed (see Fig. (\ref{current})). The gradual evanescence of transverse/shear sound propagation is mirrored by the parallel disappearance of the positive sound dispersion, a thoroughly studied phenomenon representing the most spectacular manifestation of $THz$ viscoelasticity (see Fig. {\ref{phonon}}). Such a link urges us to conclude that the physical mechanism  responsible for the positive sound dispersion is firmly associated with the shear relaxation, i.e. the relaxation involving the shear component of the viscosity. At short times (high $q$ values) over which such a shear relaxation is not yet accomplished, the fluid response resembles the one of a solid, being characterized by both
a higher sound velocity (positive sound dispersion) and the
onset of a transverse/shear propagation. Conversely, within the high-T compressed-gas region, such a solid-like (or
 elastic) response can never be probed since in such a regime the fluid has lost its viscoelastic properties. We finally observe that in all probed thermodynamic conditions the $q$-dispersion
overall bears evidence for the crossover from a low-$q$ adiabatic regime to a high-$q$ isothermal one at frequency $\sim$ 1 THz (see Fig. (\ref{phonon})).

The $q$-dispersion of the transverse acoustic (TA) mode (see Fig. (\ref{sound})) bears evidence for zones of forbidden propagation (phononic gaps). These gaps extend over the narrow-$q$ range (at T=300 K the gap expends over 0-2.6 nm$^{-1}$) and over the wider-$q$ ranges  at higher temperatures (at T=400 K and T=600 K the gaps expend over 0-7.4 nm$^{-1}$ and over the all $q$-range respectively) manifesting the inability of a fluid to support transverse/shear sound propagation, as directly predicted by Eq. (\ref{EHam}). We anticipate that the results of the present work are relevant in the field and will pave the way to future investigations of more complex molecular fluids in deeply supercritical conditions. The outcome of these studies is deemed to have a major impact onto disciplines as diverse as geophysics, planetary science and material science.

\begin{acknowledgement}
We thank Bogdan M. Leu and Ayman H. Said for their support during the experiment at Sector 30 at APS, and Sergey N. Tkachev for his help with the GSECARS gas loading system. The work at the National Synchrotron Light Source-II, Brookhaven National Laboratory, was supported by the U. S. Department of Energy, Office of Science, Office of Basic Energy Sciences, under Contract No. DE-SC00112704. Synchrotron experiment was performed at 30-ID beamline, Advanced Photon Source (APS), Argonne National Laboratory. Use of the Advanced Photon Source was supported by the U. S. Department of Energy, Office of Science, Office of Basic Energy Sciences, under Contract No. DE-AC02-06CH11357.
\end{acknowledgement}

\end{document}